\documentclass[prl,twocolumn,superscriptaddress]{revtex4-1}
\usepackage{amssymb}
\usepackage{amsfonts}
\usepackage{bbm}
\usepackage{mathrsfs}
\usepackage{graphics,graphicx,epsfig,bm,amsmath,amsthm,amssymb}
\usepackage{bm}
\usepackage{bbm}
\usepackage{longtable}
\usepackage{multirow}
\usepackage{array}
\usepackage{color}
\usepackage{cases}
\usepackage{booktabs }
\usepackage[usenames,dvipsnames]{xcolor}

\usepackage{xcolor}
\usepackage[none]{hyphenat}
\usepackage{float}
\usepackage{amsmath}
\usepackage[a4paper,colorlinks=true,
linkcolor=blue,citecolor=blue,
pdfauthor={ },
pdftitle={ },
pdfsubject={ },
pdfkeywords={ }]{hyperref}

\begin{document}
\title{Observation of a new interaction between a single spin and a moving mass}
\author{Xing Rong}
\affiliation{Hefei National Laboratory for Physical Sciences at the Microscale and Department of Modern Physics, University of Science and Technology of China, Hefei 230026, China}
\affiliation{CAS Key Laboratory of Microscale Magnetic Resonance, University of Science and Technology of China, Hefei 230026, China}
\affiliation{Synergetic Innovation Center of Quantum Information and Quantum Physics, University of Science and Technology of China, Hefei 230026, China}

\author{Man Jiao}
\affiliation{Hefei National Laboratory for Physical Sciences at the Microscale and Department of Modern Physics, University of Science and Technology of China, Hefei 230026, China}
\affiliation{CAS Key Laboratory of Microscale Magnetic Resonance, University of Science and Technology of China, Hefei 230026, China}
\affiliation{Synergetic Innovation Center of Quantum Information and Quantum Physics, University of Science and Technology of China, Hefei 230026, China}

\author{Maosen Guo}
\affiliation{Hefei National Laboratory for Physical Sciences at the Microscale and Department of Modern Physics, University of Science and Technology of China, Hefei 230026, China}
\affiliation{CAS Key Laboratory of Microscale Magnetic Resonance, University of Science and Technology of China, Hefei 230026, China}
\affiliation{Synergetic Innovation Center of Quantum Information and Quantum Physics, University of Science and Technology of China, Hefei 230026, China}

\author{Diguang Wu}
\affiliation{Hefei National Laboratory for Physical Sciences at the Microscale and Department of Modern Physics, University of Science and Technology of China, Hefei 230026, China}
\affiliation{CAS Key Laboratory of Microscale Magnetic Resonance, University of Science and Technology of China, Hefei 230026, China}
\affiliation{Synergetic Innovation Center of Quantum Information and Quantum Physics, University of Science and Technology of China, Hefei 230026, China}

\author{Jiangfeng Du}
\email{djf@ustc.edu.cn}
\affiliation{Hefei National Laboratory for Physical Sciences at the Microscale and Department of Modern Physics, University of Science and Technology of China, Hefei 230026, China}
\affiliation{CAS Key Laboratory of Microscale Magnetic Resonance, University of Science and Technology of China, Hefei 230026, China}
\affiliation{Synergetic Innovation Center of Quantum Information and Quantum Physics, University of Science and Technology of China, Hefei 230026, China}

\date{\today}

\begin{abstract}
Searching for physics beyond the standard model is crucial for understanding the mystery of the universe, such as the dark matter. We utilized a single spin in a diamond as a sensor to explore the spin-dependent interactions mediated by the axion-like particles, which are well motivated by dark matter candidates. We recorded non-zero magnetic fields exerted on the single electron spin from a moving mass. The strength of the magnetic field is proportional to the velocity of the moving mass. The dependency of the magnetic field on the distance between the spin and the moving mass has been experimentally characterized. We analyzed the possible sources of this magnetic signal, and our results provide highly suggestive of the existence of a new spin-dependent interaction. Our work opens a door for investigating the physics beyond the standard model in laboratory.
\end{abstract}

\maketitle

Spins play essential roles in nowadays science and technologies. Recently, spins in solid-state systems, most prominently the electronic spins of the nitrogen-vacancy (NV) centers in diamond, have been widely utilized as sensitive probes for several important physical quantities\cite{Degen_RMP2017,Chen2018}, such as magnetic fields. Single NV centers in diamond are defects composed of a substitutional nitrogen atom and a neighboring vacancy\cite{Gruber1997,Doherty2013}. The size of the of the NV probe can be fabricated to be as small as the nanoscale\cite{Schirhagl2013}. The geometry provides the possibility to enable close proximity between the probe and the target. Furthermore, the intrinsic noise of the probe has been well investigated\cite{Hanson2006,Maze2008,Bauch2020}. Thus this type of probe has been utilized to search for physics beyond the standard model\cite{Rajendran2017,Rong_2018,Rong_2018PRL,Man_2020}. Experiments  based on NV centers have been carried out to explore monopole-dipole interaction between the electron spin and nucleons\cite{Rong_2018},  axial-vector mediated interaction between polarized electrons\cite{Rong_2018PRL} and  an exotic parity-odd spin- and velocity- dependent interaction\cite{Man_2020}.

\begin{figure}
\centering
\includegraphics[width=1\columnwidth]{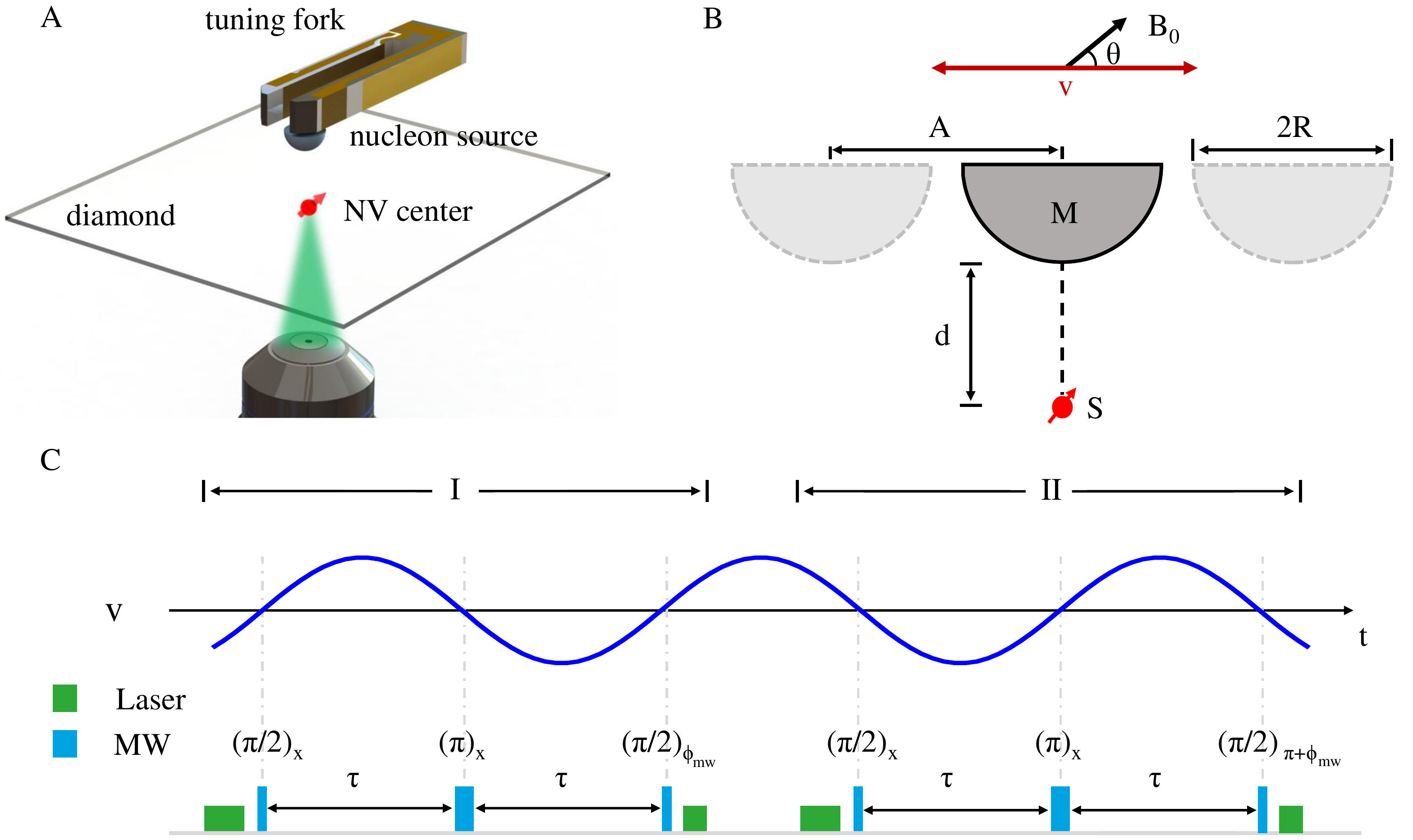}
\caption{(a) Schematic of our experimental setup. A fused silica half-ball is labeled as nucleon source, which is placed on a tuning fork of the AFM. Green laser pules have been applied for initializing and measuring the state of the NV center, which is close to the surface with depth less than 10 nm. (b) A schematic of the NV center and the moving mass. The radius of the nucleon source, M, is $R = 250 ~\mu $m. A static magnetic field $\mathbf{B_{0}} = 476~$ Gauss is applied along the symmetry axis of the NV center (S). The relative velocity vector between the NV center and nucleon half-sphere lens is labeled as $\bm v$.  The direction of $\bm v$ is parallel to the surface of the diamond.The angle between  $\bm v$ and $\mathbf{B_0}$ is $\theta =\arcsin(1/\sqrt{3})$. The amplitude of the vibration is A. The distance between the bottom of M and S is d, when S locates below the center of M.
(c) Experimental scheme of searching for exotic spin-dependent interaction between a single spin and a moving mass source.  Symbols $\pi$ and $\pi/2$ stand for the rotation angles of the quantum state due to microwave pulses. $\phi_{mw}$ ($\phi_{mw}+\pi$) is the phase of the last $\pi/2$ microwave pulse. For accumulating a positive (negative) phase factor, the $\pi/2$ microwave pulses were applied on S when M achieved the maximum value of vibration amplitude in the left (right). $\pi$ microwave pulses were applied on the middle of the microwave sequence. The time duration between $\pi/2$ and $\pi$ pulses are $\tau = \pi/\omega_M$.
}
\label{Fig1}
\end{figure}

In this paper, we conducted an experiment to search for possible exotic spin-dependent interaction which may be introduced by the moving mass. The experimental setup is a homebuilt NV-based magnetomertery combined with an atomic force microscope (AFM) as shown in the Fig.~\ref{Fig1}A. The single NV center in diamond has been taken as an atomic scale magnetometer to capture the possible magnetic signal from a moving mass, which is a fused silica half-sphere lens with diameter being 500 $\mu$m. The single NV centers were created by implantation of 10 keV $N_2^+$ ions into $<$100$>$ bulk diamond and annealing for two hours at $800~^{\circ}$C. After annealing, the diamond was oxidatively etched for 4 hours at $580~^{\circ}$C. To enhance the detection efficiency, we fabricated nanopillars on the surface of the diamond by electron beam lithography and reactive ion etching.  The typical photoluminescence rate in our experiment is 350 kcounts/s with  laser power being about $200~\mu$W. The NV center's ground state is an electronic spin triplet state $^3A_2$ with three substates $|m_S=0\rangle$ and $|m_S=\pm1\rangle$, respectively. An external magnetic field with the strength being $476~$Gauss along the symmetry axis of the NV center was utilized to remove the degeneracy of the $|m_S=\pm1\rangle$.  Two spin states, $|m_S=0\rangle$ and $|m_S=-1\rangle$, were encoded as a quantum probe. The state of the probe can be initialized and measured by green laser pulses. The evolution of the state of the probe can be engineered by microwave pulses which are delivered by a copper wire.  The source of the nucleon was a fused silica half-sphere lens installed on a tuning fork of the AFM.  Hereafter, the single electron spin of NV center and the moving mass source are denoted as S and M for convenience, respectively.

Fig.~\ref{Fig1}B shows the geometric parameters of our experiment setup. When S is positioned exactly under the center of M,  the distance between the bottom of M and S is denoted as d. The tuning fork drives the mass to move with the amplitude, A, and the frequency, $\omega_M$. The velocity of M is $v (t) = A\omega_M\sin(\omega_M t)$. An external magnetic field $\textbf{B}_0$ is applied along the NV axis. The angle between the direction of the velocity, $\bm v$, and the magnetic field, $\textbf{B}_0$, is $\theta =\arcsin (1/\sqrt{3})$ as shown in the inset of Fig.~\ref{Fig1}B. The direction of the velocity is parallel to the surface of the diamond. In the standard model, there is no mechanism, which allows possible interaction between the spin and the nucleons. However, possible interactions between spin and nucleons mediated by the axion-like particles have been proposed in the literature\cite{Moody_1984,JHighEnergyPhys_MacroForce,Fadeev2019}. In this work we conducted a  search for possible magnetic field felt by the spin from the nucleon source, which is a moving mass. The magnetic field, which is aroused from the moving mass, is denoted as $\text B_{\text {eff}}$.

Fig.~\ref{Fig1}C shows the experimental pulse sequence to search for possible exotic spin-dependent interaction.
The time evolution of  the velocity of M, $\text {v(t)}$ (blue line) has been presented in the upper panel. In the lower panel, we show the laser and the microwave pulse sequences, which were applied on S for initializing, manipulating and measuring the states.  In our experiment,  a pulse generator and a comparator were utilized to synchronize the laser and microwave pulse sequences with the vibration of M.  The first laser pulse and the following $\pi/2$ microwave pulse were performed to initial  the state of S to a superposition state $(|0\rangle-i|-1\rangle)/\sqrt{2}$. The $\pi/2$ microwave pulses were applied on S when the vibration of M achieved the maximum value of the amplitude in the left.  In this case, during the first waiting time $\tau$, the state of spin evolves around the z axis and accumulates a phase factor dependent on the magnetic field $\text B_{\text {eff}}$, while S will accumulate a phase factor with an opposite sign during the second waiting time. Because a $\pi$ pulse inverting the state of S is applied in the middle of the time evolution, the final state of S will acquire a phase factor $\phi =  \int_0 ^\tau \gamma_e \text B_{\text{eff}}(t) dt- \int_{\tau} ^{2\tau} \gamma_e \text B_{\text{eff}}(t) dt$, where $\text B_{\text{eff}}(t)$ stands for the possible magnetic field from the moving mass. After the last $\pi/2$ pulse with a variable microwave pulse phase $\phi_{mw}$ together with a laser pulse (as shown in the Part I of Fig.~\ref{Fig1}C), the population of the final state on $|m_S = 0\rangle$ is $P_+ = [1+ \cos(\phi_{mw} + \phi)]/2$.
In part II of the Fig.~\ref{Fig1}C, The final state will acquire a phase factor with the opposite sign with $\pi/2$ pulses being applied on S when M reaches the maximum value of the vibration amplitude in the right. The population of the final state on $|m_S = 0\rangle$ in this case is  $P_- = [1+  \cos(\phi_{mw}+\pi - \phi)]/2$, when the phase of the last microwave $\pi/2$ pulse is set to $\phi_{mw} + \pi$. Then we recorded the sum of the two populations, $I =P_++P_-= 1 -\sin(\phi_{mw})\sin(\phi)$.

\begin{figure}[http]
\centering
\includegraphics[width=\columnwidth]{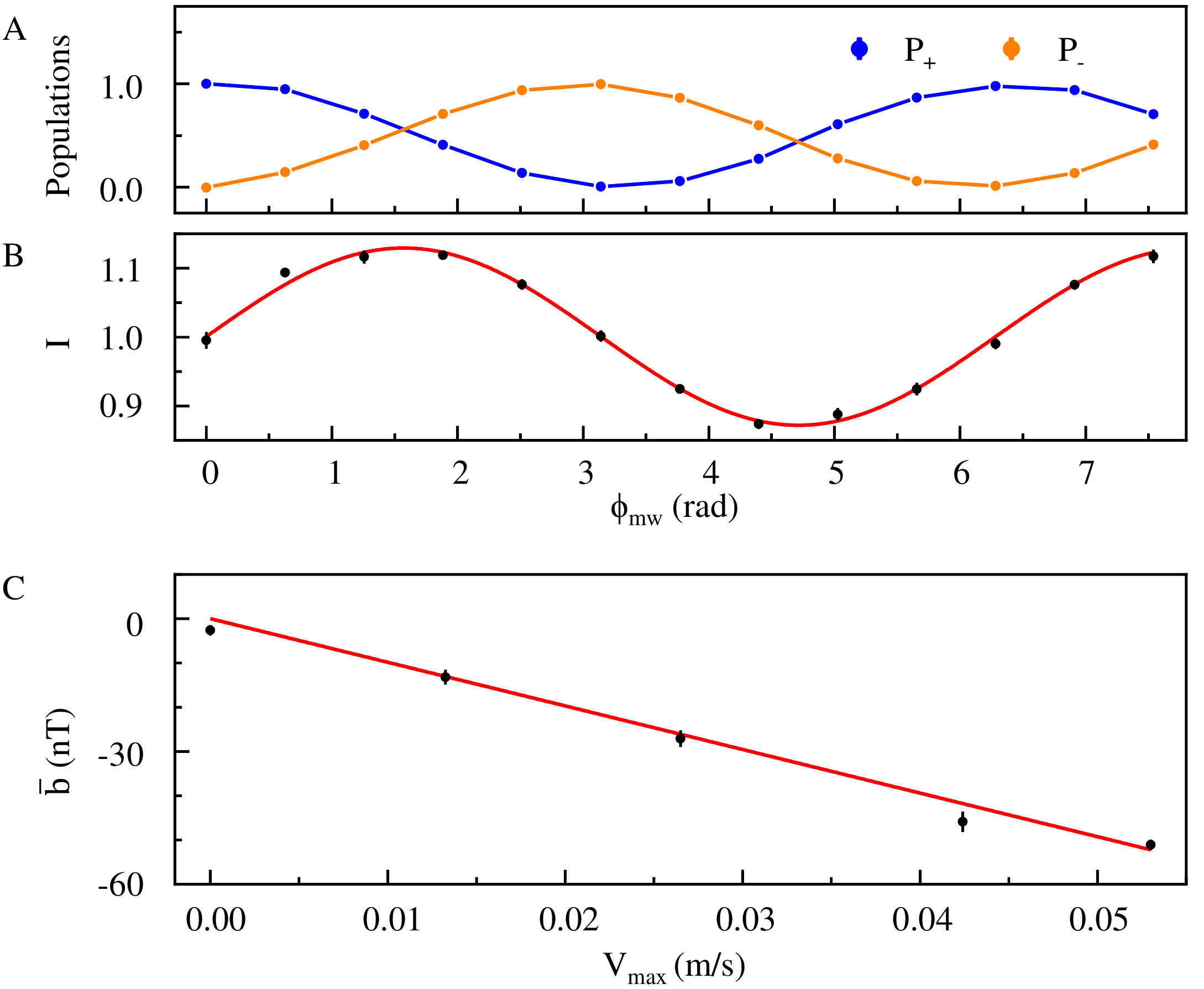}
    \caption{Experimental observation of nonzero magnetic field due to the moving mass. (a) shows experimental data for  populations of the final state on $|m_S = 0\rangle$,  $P_+ $ (blue dots with error bars) and  $P_-$ (orange dots with error bars), respectively. (b) shows the experimental data of the sum of the two populations. The experimental data (black dots with error bars) were fitted to $I = 1 -\sin(\phi_{mw})\sin(\phi)$. The experimental result shows that there is a nonzero phase factor $\phi = -0.128(3)$. The average magnetic field $\overline{b} = -51(1)~$nT can be obtained.   (c) The dependency of the magnetic field on the velocity of the mass. The x-axis is the maximum value of the velocity of the mass, $v_{max}$. Black dots with error bars are experimental data of the magnetic field, $\overline{b}$, which were fitted by a liner function $\overline{b}  = k v_{max}$, with the slope $k = (-985\pm30)~$ (nT $\cdot$ s/m).
 }
    \label{Fig2}
\end{figure}

In our experiment, the waiting time duration is $\tau = 6.994~\mu$s, the amplitude of the vibrating M is $A = 120~$nm, and the angular frequency $\omega_M = 2\pi \times 70.47~$kHz. The distance between the NV center and the bottom of the half-sphere was tuned from $1~\mu$m to $12~\mu$m. The decoherence time of this NV center was $T_2 = 77(3)~\mu$s measured by a spin echo experiment. The pulse lengths of $\pi/2$ and $\pi$ are $102~$ns and $205~$ns, respectively. The time durations of laser pulses for initialization and readout are $2.0~\mu$s and $0.3~\mu$s, respectively.  Fig.~\ref{Fig2}  shows the experiment results when the distance was set to be $1~\mu$m. We repeated the measurement for 8 millions times to build good statistics. Fig.~\ref{Fig2} A shows the experimental results of population of the final state on $|m_S = 0\rangle$ , $P_+$ (blue dots with error bars) and  $P_-$ (orange dots with error bars), respectively. Fig.~\ref{Fig2} B is experimental data of the sum of the two populations. From Fig.~\ref{Fig2} B, a nonzero phase factor, $\phi = -0.128(3)~$rad, has been observed in our experiment. An average magnetic field $\overline{b} = -51(1)~$nT during the evolution time $\tau$ can be obtained by $\overline{b} = \phi/2\gamma_e\tau$, where $\gamma_e$ is the gyromagnetic ratio of the electron spin.

We now analyze the possible sources of this nonzero magnetic field. The modulated magnetic field due to the diamagnetism of M should be taken into account. Since  a $\pi$ pulse  was applied in the middle of the pulse sequence, the magnetic field  due to  diamagnetism of M is $|B_{diam}|<2.3\times 10^{-7}~$nT when d is $1~\mu$m.  If there is electric charge in the half-sphere, a magnetic field introduced by the moving charge should also be taken into account. First we estimated the quantity of the electric charge on the half-sphere.  We measured the resonance frequencies of the transition between $(|0\rangle$  and $|-1\rangle)$ when the half-sphere was placed near and far away from the NV center. The electric field due to the charge will shift the resonance frequency. In our experiment, the frequency shift is smaller than $0.1~$MHz, and the maximum possible charge on the half-sphere is estimated to be $5\times 10^{-14}$C, when the screening effect due to the surface of the diamond has been taken into account\cite{Oberg2020}. Thus the magnetic field due to this effect is $|B_{Q}|< 2~$nT when d is $1~\mu$m. Both B$_{diam}$ and B$_{Q}$ are at least one order of  magnitude smaller than the observed magnetic signal when d is $1~\mu$m. The detailed analysis is included in the Appendix B.

A possible source of the nonzero magnetic field observed in our experiment can be attributed to a type of exotic parity-even spin- and velocity-dependent interaction\cite{JHighEnergyPhys_MacroForce}, which is described by an effective potential
\begin{equation}
V = f^{\perp}\frac{\hbar^2}{8 \pi m_ec}\bm{\sigma} \cdot \bm{v}\times \hat{r}( \frac{1}{\lambda r} + \frac{1}{r^2}) e^{{-\frac{r}{\lambda}}},
\label{Vexotic}
\end{equation}
where $\bm{\sigma}$ is Pauli vector of the electron spin, $r=|\textbf{r}|$ with $\textbf{r}$ being the displacement vector between the electron and nucleon,  $\hat{r}$ is the unit vector between the electron spin and nucleon, $\bm{v}$ is their relative velocity, $\lambda=\hbar/(m_\textrm{b}c)$  is the force range with $m_\textrm{b}$ being the boson mass, c being the speed of light in vacuum, $m_e$ is the mass of the electron and $\hbar$ being the Planck's constant divided by $2\pi$. Here $f^{\perp}$ is the dimensionless coupling constant\cite{Leslie2014}. This coupling constant may be the combination of the scalar electron coupling and the scalar nucleon coupling for spin-0 exchange. It could also be the vector electron coupling and the vector nucleon coupling for spin-1 exchange\cite{JHighEnergyPhys_MacroForce,Leslie2014}. Recently, experimental searching for this interaction have been performed by several laboratory experiments\cite{Ding2020,Kim2018,Piegsa2012,Haddock2018,Heckel2006,Heckel2008}.
This type of interaction leads to an effective magnetic field felt by the electron spin arising from the moving nucleons,
\begin{equation}
B(r)=f^\perp\frac{\hbar}{4\pi m_ec\gamma_e}v_y\frac{(-r_x\sin{\theta}+r_z\cos{\theta})}{r}(\frac{1}{\lambda r}+\frac{1}{r^2})e^{-r/\lambda}
\label{Beff}
\end{equation}
where $\gamma_e$ is the gyromagnetic ratio of the electron spin, $v_y$ is the velocity of the moving nucleon, $r_x$ and $r_z$ are the x and z components of $\bm{r}$.
Since the nucleon source in our experiment is a half-ball, the magnetic field arising from this hypothetic spin- and velocity-dependent interaction can be derived from integrating $\textbf{B}(r)$ over all nucleons of M as
\begin{equation}
\textbf{B}_{\text{eff}}= \int_V \rho\textbf{B}(r) dV.
\end{equation}
The number density of nucleons in M in our experiment is $\rho=1.33\times10^{30}~$m$^{-3}$.

We performed an experiment to determine the dependency of the magnetic field on the velocity. Fig.~\ref{Fig2}C show the experimental data about the dependency of the magnetic field on the velocity when the distance, d, was set to be $1~\mu$m. The x-axis is the maximum value of the velocity of during the vibration and the y-axis is the measured average magnetic field. It is clear that the magnetic field is proportional to the velocity of the moving mass.

\begin{figure}[http]
\centering
\includegraphics[width=\columnwidth]{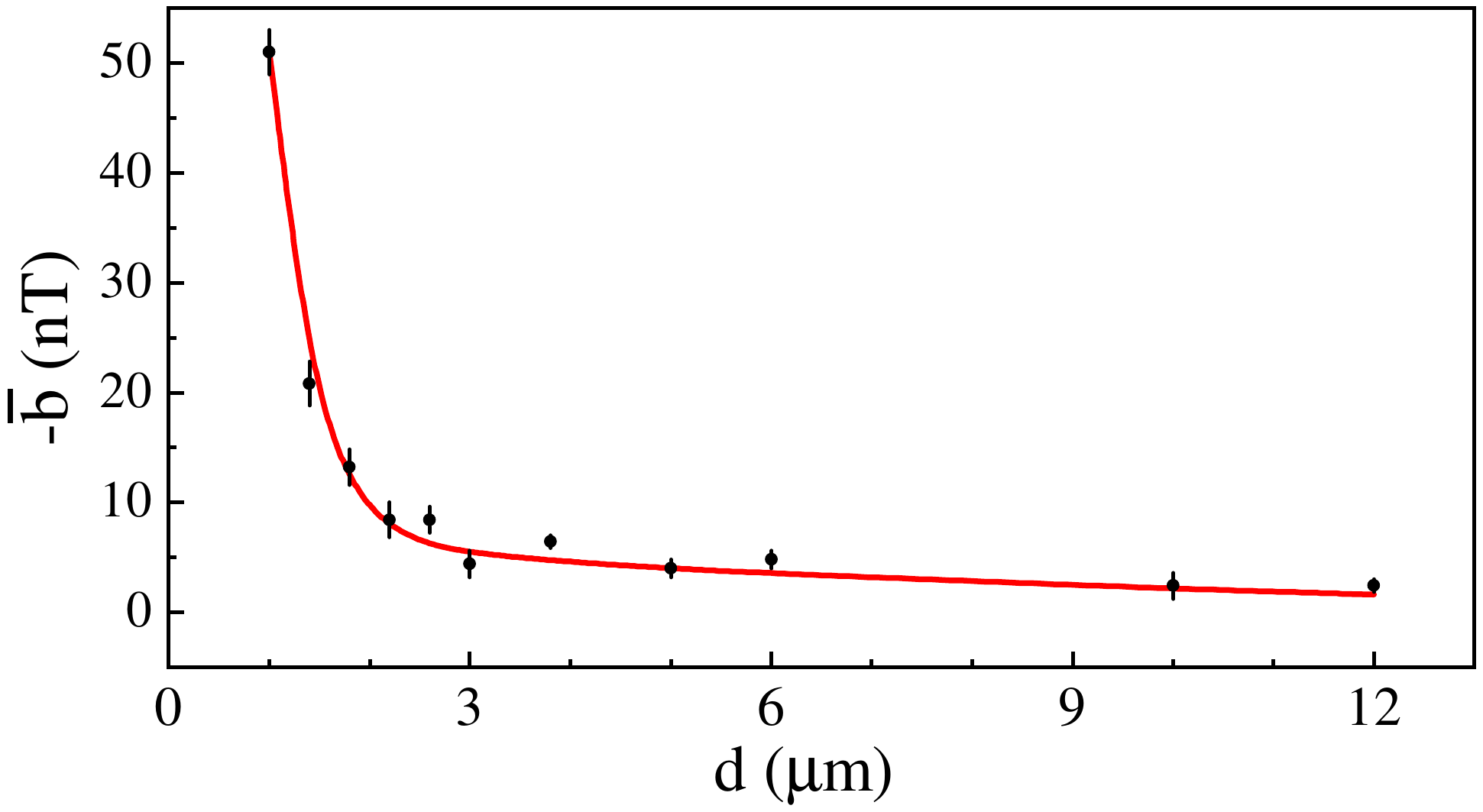}
    \caption{Dependence of the experimental measured magnetic field, $\overline{b}$, on the distance, d. Black Dots with error bars are experimental values of the magnetic field on the S from the moving mass source M. Experimental data are fitted by equation \ref{fit-function}.
 }
    \label{Fig3}
\end{figure}

Fig.~\ref{Fig3} shows the experimental data of the dependency of the magnetic field, $\overline{b}$, on the distance $d$. The distance has been tuned from $1~\mu$m to $12~\mu$m in this experiment, when the maximum value of the velocity was fixed to be $0.053~$m/s. We found that there is a fast decay region when $d<3~\mu$m, which indicates a small force range. When $d>3~\mu$m, the decay of the magnetic signal is much slower, which indicates a much larger force range. Therefore, we utilized the following function to fit the experimental data,
\begin{equation}
\overline{b}=\frac{1}{\tau}\int_0^\tau \left( \sum_{i=1}^{200} B_{eff}\left( \lambda_i, f ^{\perp}\left( \lambda_i \right),t \right) \right) dt ,
\label{fit-function}
\end{equation}
where $f (\lambda)= f ^{\perp}_1 g(\lambda, \lambda_1,\Gamma_1) +  f ^{\perp}_2 g(\lambda, \lambda_2,\Gamma_2)$,  and the distribution function of the force range is $g(\lambda,\lambda_c,\Gamma) = \exp {[-(\lambda-\lambda_c)^2/2\Gamma^2]}$ with $\Gamma$ being  the width and $\lambda_c$ being the mean value.
 By fitting the experimental data to this function, we find that $\lambda_1 =3.82\times 10^{-7}$m with $f^{\perp}_1 = 4.83\times 10^{-6}$and $\Gamma_1 = 5\times 10^{-8}$m, and  $\lambda_2 =8.07\times 10^{-6}$m with  $f^{\perp}_2 = 3.93\times 10^{-8}$ and $\Gamma_2 = 5\times 10^{-8}$m.

\begin{figure}[http]
\centering
\includegraphics[width=\columnwidth]{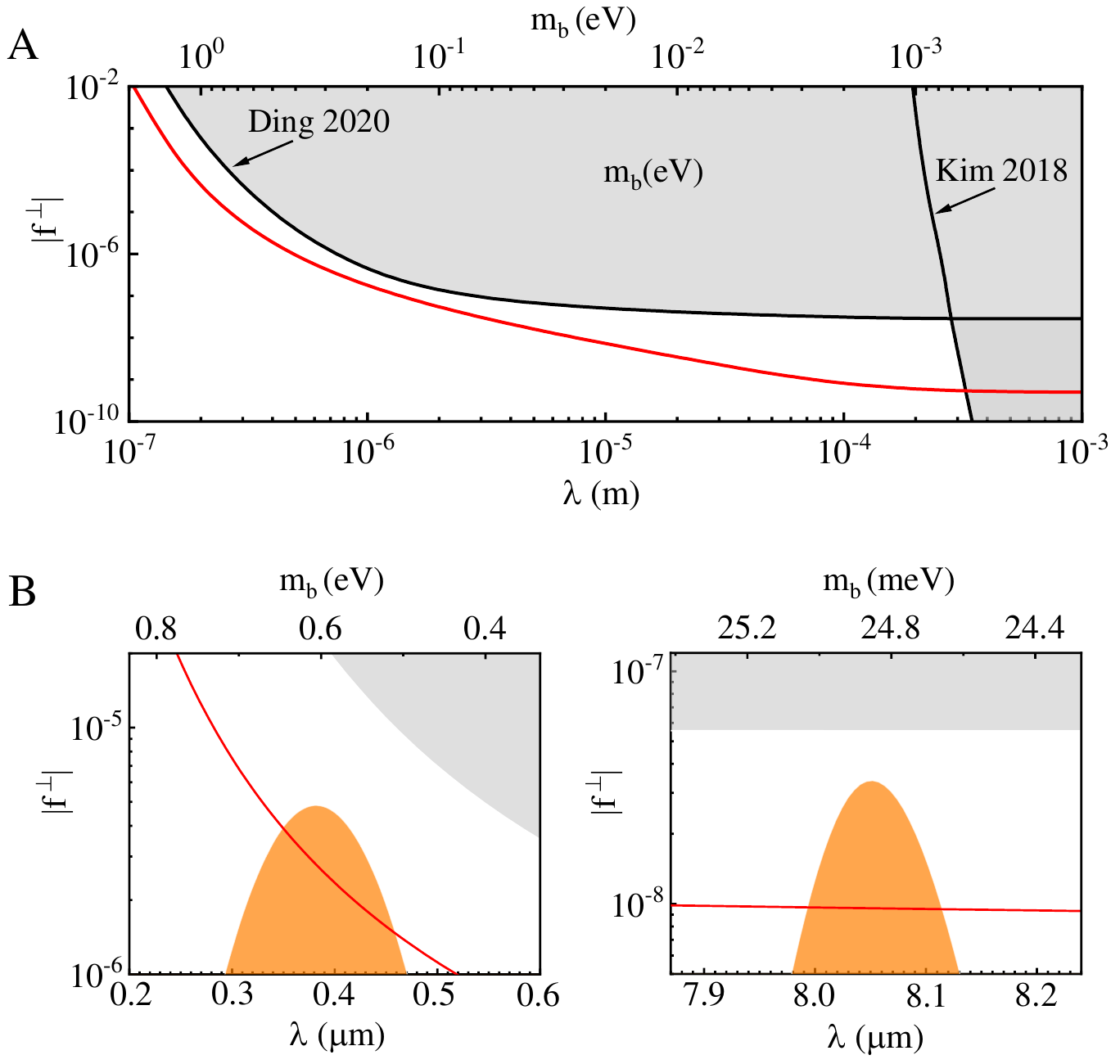}
    \caption{(a) Red line is the experimental precision in this work, while black lines are constraints established by the previous experiments. (b) the experimental spectrum of the new bosons. Grey parts are regions excluded by previous experiment\cite{Ding2020}. Red lines are the precision of our experiment. Orange regions are the distributions of the force ranges and the corresponding coupling strengths.
 }
    \label{Fig4}
\end{figure}

In Fig.~\ref{Fig4}, the experimental spectrum of the possible new bosons which mediate the exotic parity-even spin- and velocity-dependent interaction has been presented. We first compare the experimental precision obtained from the noise level of our experiment with previous experiments\cite{Ding2020,Kim2018} in Fig.~\ref{Fig4}A. In the region $\lambda < 300 ~\mu$m, our results provide better experimental precision than literatures. The precision of our experiment for the force range $\lambda = 200~\mu$m is about two orders of magnitude better than previous experiments.  In Fig.~\ref{Fig4}B, the grey areas are the region excluded by previous experiment\cite{Ding2020} and  orange areas are the spectrum of the new bosons obtained from our experiment. There are two main peaks in the spectrum. One peak is at the shorter force range with the mean force range being $\lambda_1 =3.82\times 10^{-7}$m. The other peak sits at the longer force range with the mean force range being $\lambda_2 =8.07\times 10^{-6}$m. Both peaks are below the constraints set by literatures and stronger than the precision of our experiment.

Discussion. We  have observed nonzero magnetic fields on a single electron spin from a moving source. Possible sources of the magnetic signal, such as diamagnetism of M and moving charge of M have been carefully analyzed. The magnetic signal from these effects are much weaker than what we have observed. A possible explanation of such magnetic signal could be due to a type of parity-even spin- and velocity- dependent interaction.  Our results agree with this type of the interaction well. The force range and the corresponding coupling strength were obtained from our experimental data.
Our results provide highly suggestive of the evidence for the new spin-dependent interaction. In future, we expect that more detailed experiments will be carried out to investigate this exotic interaction.

We thank Prof. Cai and Hang Liang for helpful discussion. This work was supported by the National Key R$\&$D Program of China (Grant Nos. 2018YFA0306600 and No. 2016YFB0501603), the National Natural Science Foundation of China (Grants Nos. 11722327, 11961131007 and 11653002), the Chinese Academy of Sciences (Grants No. GJJSTD20200001, No. QYZDY-SSW-SLH004 and No. QYZDB-SSW-SLH005), and Anhui Initiative in Quantum Information Technologies (Grant No. AHY050000).
X.\ R thank the Youth Innovation Promotion Association of Chinese Academy of Sciences for the support.

Xing Rong, Man Jiao and Maosen Guo contributed equally to this work.

\appendix
\section {Experimental setup}
\label{appendix.A}
\begin{figure}[http]
\centering
\includegraphics[width=0.75\columnwidth]{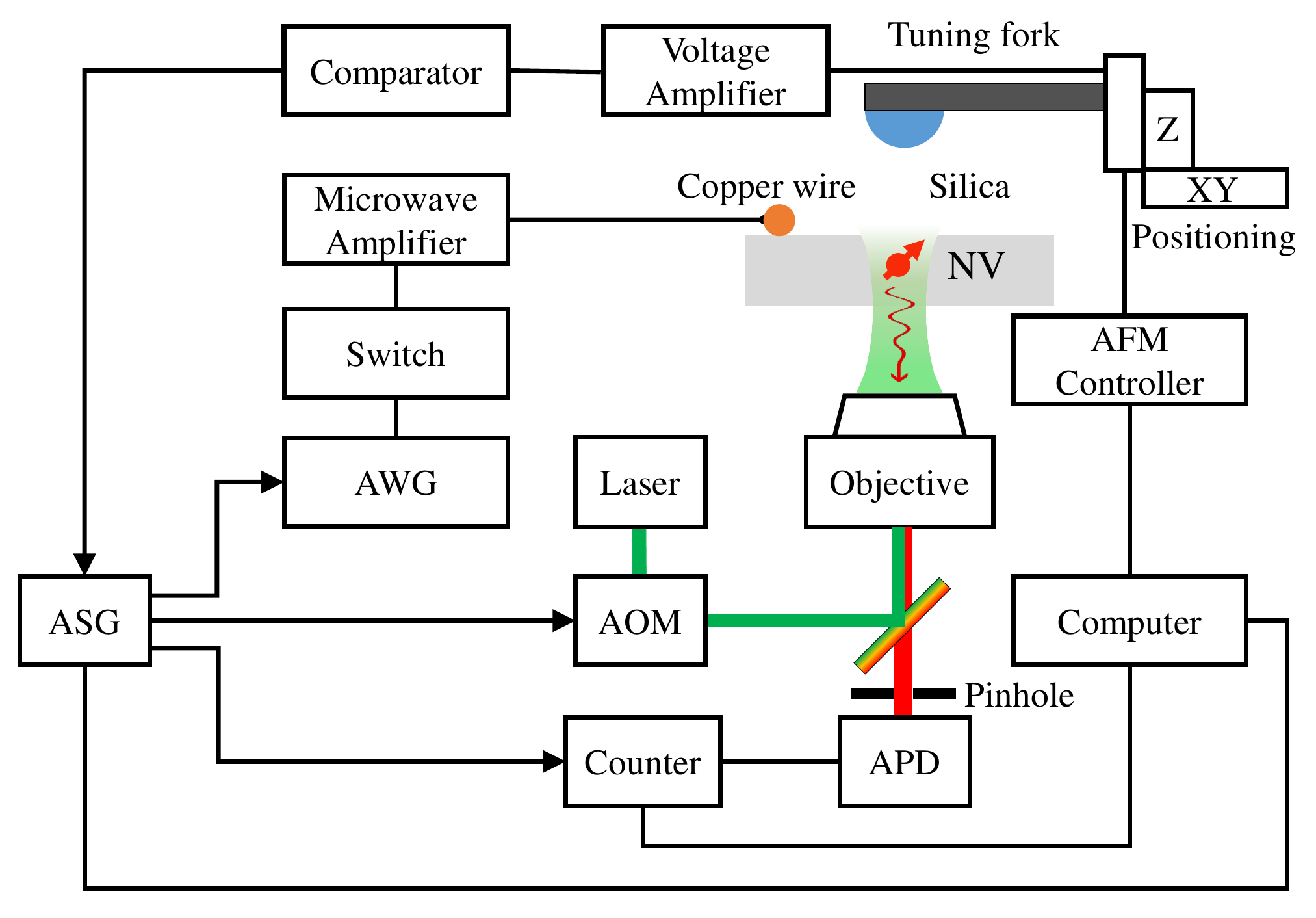}
    \caption{\textbf{Schematic of the experimental setup.}
}
    \label{FigS1}
\end{figure}

To search for the spin-dependent interaction, we utilized an optically detected magnetic resonance (ODMR) setup together with an atomic force microscope (AFM). The schematic diagram of the experimental setup is shown in  Fig. \ref{FigS1}.

The ODMR setup consists of  a confocal microscopy and a microwave system. The laser with the wavelength being 532 nm passes twice through an acousto-optic modulator (AOM, ISOMET 1250C) and an objective (Olympus, LUCPLFLN 60X; numerical aperture (NA), 0.7) before being shed on the NV center. With the pulsed laser, NV canter can be efficiently polarized and exhibit a spin-dependent photoluminescence (PL). An avalanche photodiode (APD, Perkin Elmer SPCM-AQRH-14) with a counter card is utilized to  collect the red fluorescence for the readout the state of the electron spin of NV centers. A light-emitting diode (LED) illuminator with a 470-nm wavelength (Thorlabs M470L3-C1) and a charge-coupled device (CCD) camera are used for wide-field and Newton rings views. The microwave (MW) pulses for manipulating the electron spin in NV centers,  is generated by a 12 GSa/s arbitrary waveform generator (AWG Agilent, M8190A), amplified by a 30W microwave power amplifier (Mini-Circuits, ZHL-30W-252 S+), and then delivered via a copper wire with the diameter being 15.2 $\mu$m.  An arbitrary sequence generator (ASG) \cite{Qin2016} was utilized to synchronize the experimental setup.

A quartz tuning fork has been utilized in our experiment.  A $500~\mu$m diameter silicon oxide (Edmund Optics Inc) is attached on the tuning fork. The time-varying charge induced by the vibration of the tuning fork prongs are translated to voltage with an amplifier (Mini-Circuits, HVA-10M-60-F). This AFM based on a tuning fork is controlled by a commercial controller (Asylum Research MFP3D) and mounted with a homebuilt AFM probe head. A translation stage based on a piezo motor (Physik Instrumente Q-522) is utilized for the coarse approach. Each AFM probe head and the diamond base plate contained a three-axis tilt platform driven by a stepper motor actuator (Thorlabs ZFS25B) for the accurate tilting adjustment. The whole probe head is placed in a temperature-stabilized chamber, so that the positioning thermal drift can be suppressed. The oscillation amplitude of the quartz tuning fork can be obtained from the measurement of mechanical properties of fork (spring constant, Q factor, and resonant frequency) and the current due to piezoelectricity of the tuning fork\cite{Liu2008,Dagdeviren2019}. The oscillation amplitude is $A=120~$nm in our experiment. The drive angular frequency was set to be $\omega_{M}=2\pi\times70.47~$kHz to match a natural frequency of the tuning fork. The waiting time in the microwave pulse sequences were set to $\tau = 6.994~\mu$s, accordingly. Note that the direction of the vibration of the tuning fork has been carefully set to be parallel to the surface of the diamond.

\begin{figure}[http]
\centering
\includegraphics[width=0.75\columnwidth]{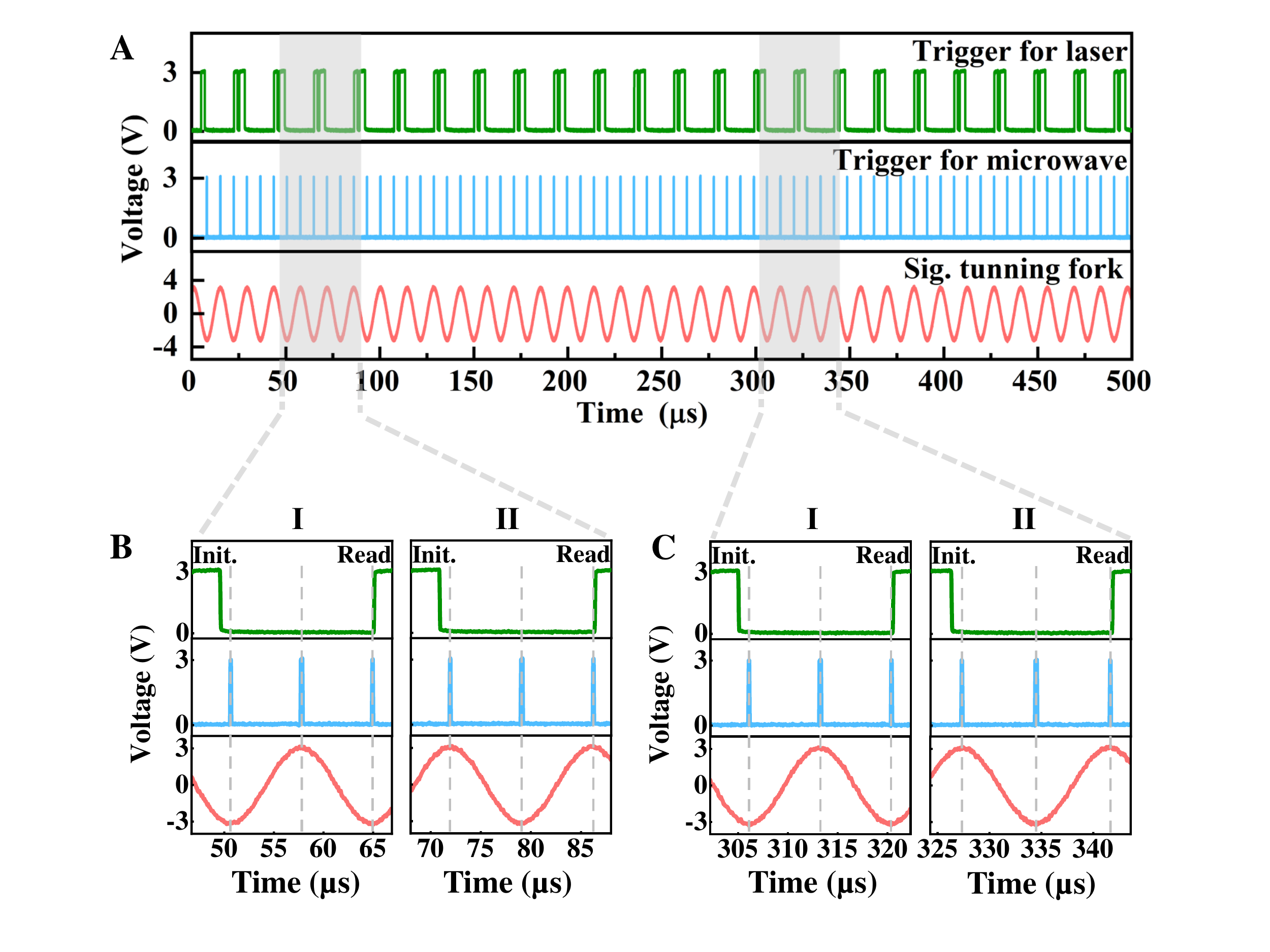}
    \caption{Experimental pulse sequences of our experiment.
}
    \label{FigS2}
\end{figure}

We utilized a a pulse generator and a comparator to synchronize the pulse sequences of the laser and microwave with the vibration of M.  In  Fig. \ref{FigS2}, we show that the microwave/laser pulse sequences and the moving of the half-sphere lens can be synchronized very well with good time precision.

\begin{figure}
\centering
\includegraphics[width=0.6\columnwidth]{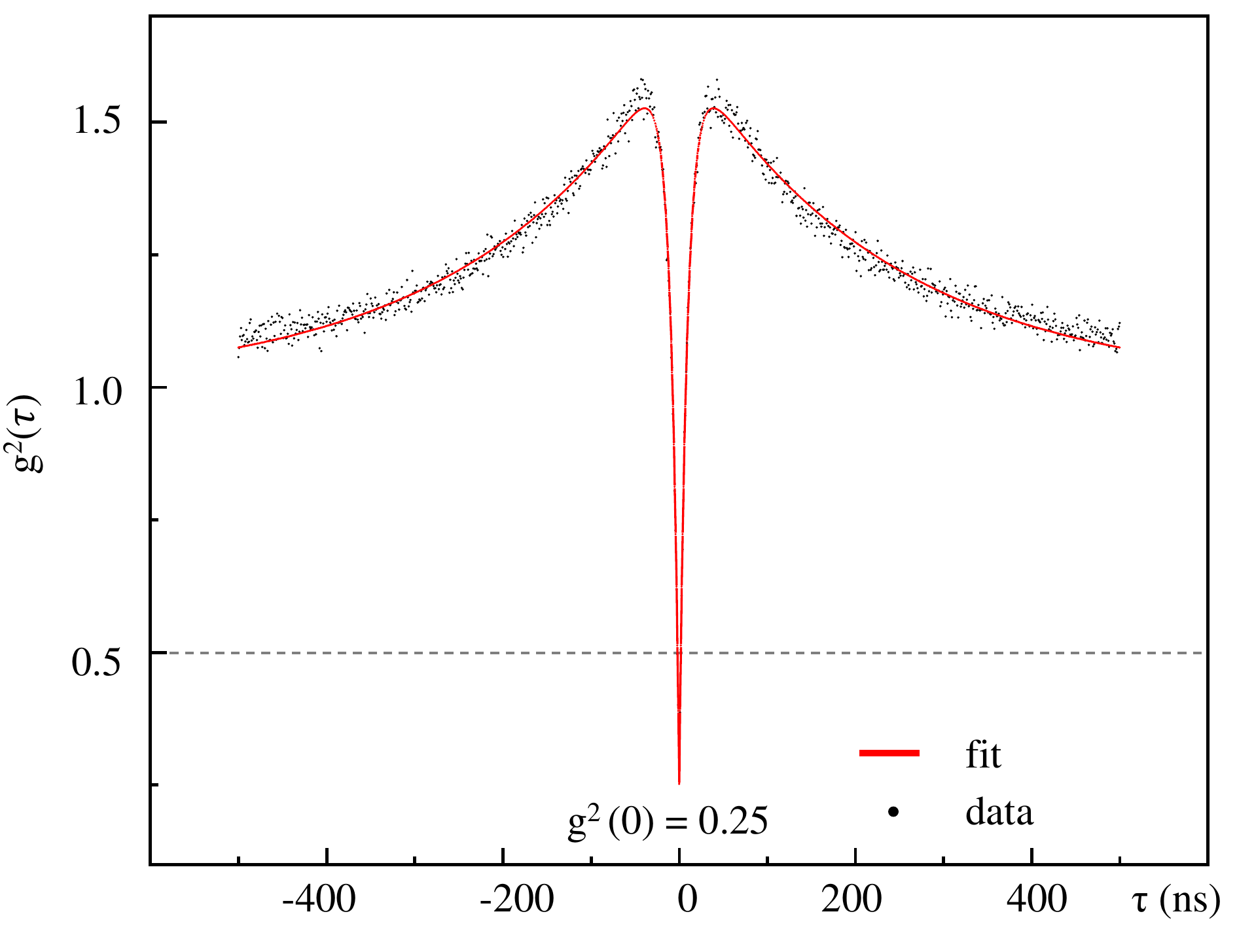}
\caption{Measurement of the second-order correlation function $g^2(\tau)$ of the NV center in this experiment with $g^2(0)= 0.25$.
    The dashed line indicates $g^2(0)= 0.5$. Black points indicate data (without background subtraction), and the red line is a fit to $g^{(2)}(\tau)=1+\frac{1}{N}\left(C_{1} \cdot e^{-\left|\tau / \tau_{1}\right|}+C_{2} \cdot e^{-\left|\tau / \tau_{2}\right|}\right)$
}
\label{FigS3}
\end{figure}

In our experiment, the external static magnetic field is $476~$G. The nuclear spin $^{14}$N can be polarized to the state $\left|+1\right\rangle_{n}$ under such condition with laser pumping\cite{Jacques2009}. Considering the subspace composed of $\left|m_{s}=0\right\rangle$ and $\left|m_{s}=-1\right\rangle$, the Hamiltonian of the NV center is
\begin{equation}
\label{Hamiltonian}
H_{NV}=DS_{z}^{2}+\gamma_{e}B_{0}S_{z}+\sqrt{2}\gamma_{e}S_{x}B_{1}\cos\left(\omega t-\varphi_{m w}\right)+\gamma_{e}S_{z}B_{eff}
\end{equation}
where $S_{x}$, $S_{z}$ are spin operators of NV center with $\mathrm{S}=\frac{1}{2}$. $D=2\pi\times2.87~$GHz is the zero field splitting. $\gamma_{e}$ is the gyromagnetic ratio of the electron. $B_{0}$ is the external static magnetic field.
$B_{1}$, $\omega_{1}$, and $\varphi_{mw}$ corresponds to the amplitude, angular frequency, and phase of the microwave pulse. $B_{eff}$ is the effective magnetic field along the NV symmetry axis due to the possible spin-dependent interaction between the electron spin and the half-ball lens.

The decoherence time of the electron spin under $476~$G magnetic field was measured by spin echo experiment \cite{PR_Echo} and the fitted coherence time is $T_2=77\pm3~ \mu$s .

\begin{figure}
\centering
\includegraphics[width=0.6\columnwidth]{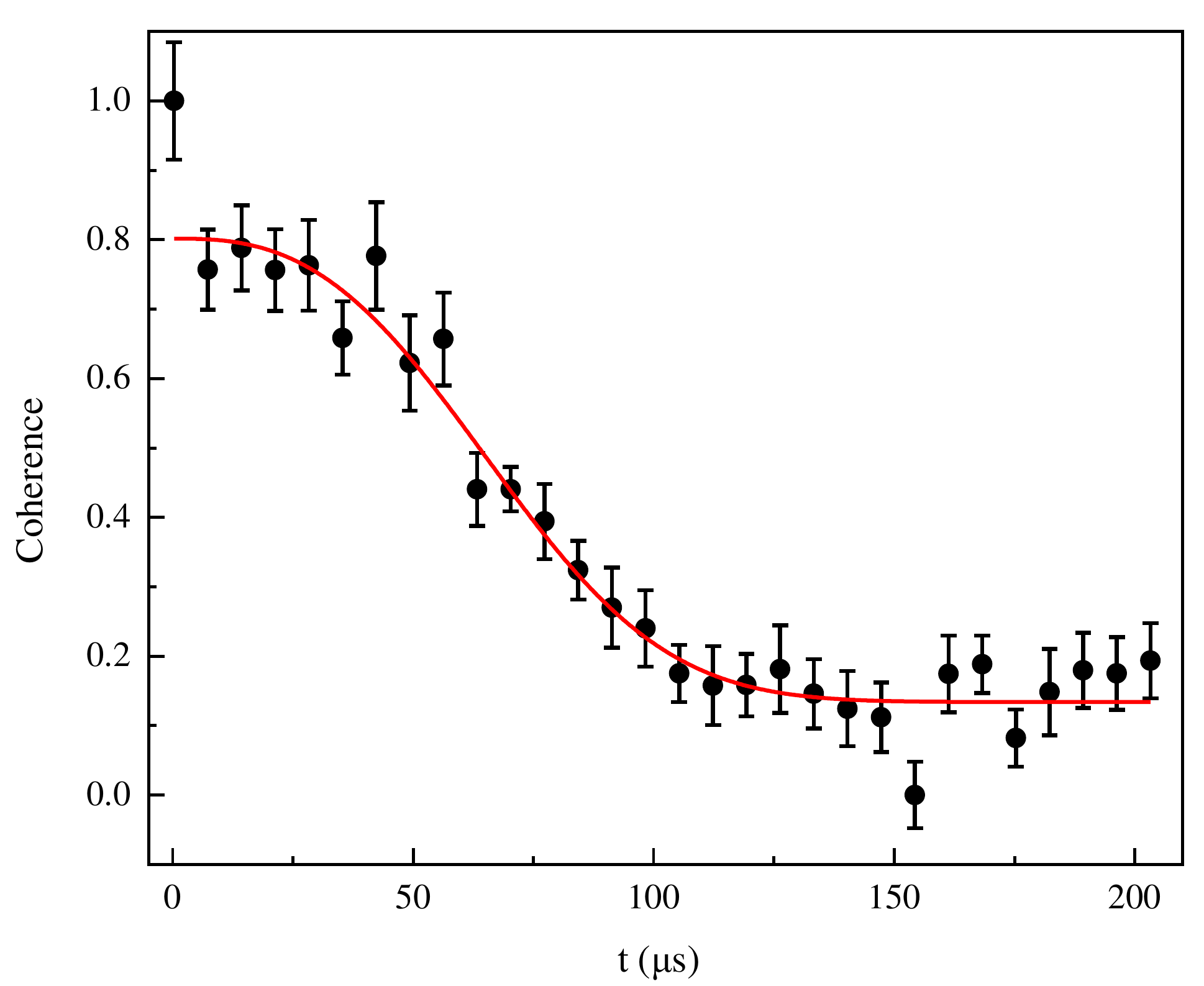}
\caption{Coherence time of the NV center using spin echo sequence.}
\label{FigS4}
\end{figure}

\section { Analysis of the possible sources of the magnetic signal}
\label{appendix.B}

\textbf{Magnetic signal due to the diamagnetism of the half sphere and tuning fork}

In the external magnetic field of $B_0=476~$Gauss, the diamagnetism of half-sphere lens and tuning fork also causes a magnetic field $B_{diam}$ on the NV center.
The magnetic field $B_{diam}$ due to the diamagnetism can be obtained by integrating over the whole volume of the sphere and tuning fork.
\begin{equation}
\mathbf{B}_{\mathrm{diam}}=\int_{V} \frac{\chi}{4 \pi}\left[\frac{3 \mathbf{r}\left(\mathbf{B}_{0} \cdot \mathbf{r}\right)}{r^{5}}-\frac{\mathbf{B}_{0}}{r^{3}}\right] d V
\label{B_diam}
\end{equation}
where $\chi=-11.28\times10^{-6}$, is magnetic susceptibility of the half sphere and tuning fork being made of SiO$_2$.

\begin{figure}
\centering
\includegraphics[width=1\columnwidth]{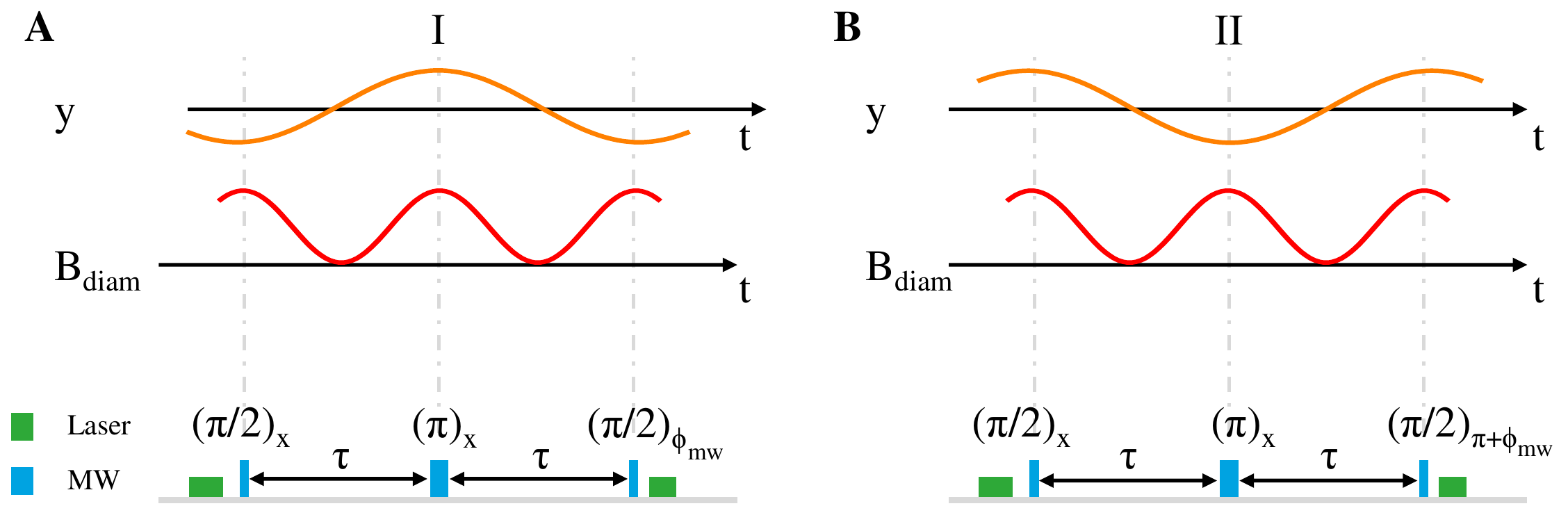}
\caption{ Vibration of the half-sphere lens, $\mathbf{B}_{\mathrm{diam}}$ due to the diamagnetism of the SiO$_2$ and the pulse sequences. The effect due to the diamagnetism can be cancelled by the pulse sequences.}
\label{FigS5}
\end{figure}

 Fig.\ref{FigS5} shows the vibration of the half-sphere lens in y axis, $\mathbf{B}_{\mathrm{diam}}$ parallel to the NV axis and the synchronized pulse sequences. Since the vibration of half-sphere lens in the first and the second $\tau$ is symmetric, the phase factor due to the $B_{diam}$ can be canceled by the Hahn-echo sequence\cite{PR_Echo}.
In addition, considering the imperfect synchronization of the Hahn-echo sequence, the difference between the first and the second waiting time in the Hahn Echo sequence is measured to be less than 40 ns. The non-zero phase factor  on the final state due to this effect is estimated to be less than
$2.9\times10^{-10}rad$, and the mean anagenetic field due diamagnetism is less than $2.3\times10^{-7}nT$ with $d=1\mu$m, which is far less than the experimental result.

Each prong of the tuning fork is a 2.36 mm $\times$ 0.59 mm $\times$ 0.33 mm tube. The distance between the undersurface of the tuning fork prong and the NV center is lager than 250 $\mu$m.  The material of the tuning fork is SiO$_2$. Because the amplitude of the vibration of the tuning fork is 120 nm, which is much smaller compared with its size and the distance. The effect due to the tuning fork is  negligible\cite{Rong_2018}.

\textbf{Magnetic signal due to the electric charge in the moving mass}

The electric charge on the half-sphere will cause the magnetic signal if the half-sphere is moving. We estimated this effect according to the following procedure.
If there is a point charge on the bottom of the half-sphere, the electric field generated by this point charge will shift of the energy lever of the electron spin of NV center. The Hamiltonian which described the coupling between the NV center and the electric field is $H_E = d_{\parallel}E_z S_z^2$ and the axial electric dipole moment of NV center is $ d_{||}=(0.35\pm0.02) $Hz cm/V \cite{Oort1990}. The resonance frequency difference $\Delta f$, when d is set from 1 $\mu$m to $20\mu$m, was $\Delta f\leq(74\pm9)~$kHz in our experiment. Then the maximum electric field felt by NV center is $\Delta E_z=\Delta f/d_{\parallel}=2.1\times10^7~$V/m. When the screening effect of the surface of the diamond\cite{Oberg2020} is considered, the change of the electric field due to the point charge is $\Delta E < 4.2\times 10^8$ V/m and the threshold of the charge is smaller than $5\times 10^{-14}$C. The average magnetic field generated by this point charge should be $|b_Q| < 2~$nT with $d=1\mu$m, which is one order of magnitude smaller than what we have observed.

\end{document}